# Which Neural Network Architecture matches Human Behavior in Artificial Grammar Learning?

Andrea Alamia, Victor Gauducheau, Dimitri Paisios, Rufin VanRullen

CerCo, CNRS, Université de Toulouse; Toulouse, 31055 (France)

## Abstract

In recent years artificial neural networks achieved performance close to or better than humans in several domains: tasks that were previously human prerogatives, such as language processing, have witnessed remarkable improvements in state of the art models. One advantage of this technological boost is to facilitate comparison between different neural networks and human performance, in order to deepen our understanding of human cognition. Here, we investigate which neural network architecture (feed-forward vs. recurrent) matches human behavior in artificial grammar learning, a crucial aspect of language acquisition. Prior experimental studies proved that artificial grammars can be learnt by human subjects after little exposure and often without explicit knowledge of the underlying rules. We tested four grammars with different complexity levels both in humans and in feedforward and recurrent networks. Our results show that both architectures can "learn" (via error back-propagation) the grammars after the same number of training sequences as humans do, but recurrent networks perform closer to humans than feedforward ones, irrespective of the grammar complexity level. Moreover, similar to visual processing, in which feedforward and recurrent architectures have been related to unconscious and conscious processes, our results suggest that explicit learning is best modeled by recurrent architectures, whereas feedforward networks better capture the dynamics involved in implicit learning.

**Keywords**: **artificial grammar learning, Chomsky hierarchy, feedforward and recurrent neural networks, recursion, implicit learning**

## Highlights

- Both feedforward and recurrent networks can learn artificial grammars after relatively few examples.
- Recurrent networks model human artificial grammar learning better than feedforward architectures.
- Feedforward and recurrent architectures better model implicit and explicit learning respectively.

# 1. Introduction

In recent years the field of neural networks has undergone a substantial revolution boosted by deep learning approaches (Lecun, Bengio, & Hinton, 2015). Different architectures have reached human-like performance in domains that were previously considered as sole prerogative of the human brain, such as perception (VanRullen, 2017) or language (Young, Hazarika, Poria, & Cambria, 2018). Part of this success originates from insights provided by cognitive sciences, in which brain-inspired solutions are implemented in functional models (Hassabis, Kumaran, Summerfield, & Botvinick, 2017). Conversely, it is possible to investigate the computational processes that take place in the human brain by comparing them with artificial functional models (Yamins & DiCarlo, 2016). For this purpose, Artificial Grammar Learning represents an ideal venue, given its well-established roots in both the cognitive and computer science literature. On the one hand, a formal definition of grammar complexity (i.e. Chomsky's hierarchy, Chomsky, 1956) provides a theoretical framework to study grammar learning; on the other hand, previous studies in humans set a well-defined experimental framework to compare human behavior with the performance of different neural network architectures.

### 1.1. Formal language theory and Chomsky's hierarchy

Formal Language Theory (FLT) stemmed from studies grounded in mathematics and computability theory, realized in the first half of the previous century by logicians such as Emil Post or Alan Turing (Post,

1944; Turing, 1937). In FLT, a *language* is defined as an infinite set of *strings*, which is in turn a sequence of symbols. Whether a string belongs or not to a language is determined by a *grammar*, i.e. a set of rules, and the distinction between grammatical and non-grammatical sequences is categorical (see Figure 1A for examples). Noam Chomsky was the first to introduce a hierarchical classification of grammars composed of 4 nested levels, sorted on the basis of computational complexity (Chomsky, 1956, 1959). The lowest level of the scale corresponds to *regular expressions*, which can be correctly classified by a memory-less finite-state automaton; the highest level coincides with the *enumerable languages*, correctly classified by Turing machines. In between, *context free* (CF) and *context specific* (CS) grammars are respectively solved by push-down and linear-bounded automata (Tecumseh Fitch & Friederici, 2012). However, this 4-levels ranking falls short when applied to natural languages (e.g. English). Consequently, one intermediate level has been added to the original hierarchy between CF and CS grammars, namely the *mildly context sensitive* grammar (Joshi, 1985), which supposedly includes all natural languages (Jäger & Rogers, 2012). However, experimental results have shown that participants can learn artificial grammars equally well irrespective of their level in the Chomsky hierarchy. This demonstrates that the hierarchy may reflect some form of computational complexity, but it does not reflect cognitive complexity (Öttl, Jäger, & Kaup, 2015); possibly, different computational processes may be involved in human grammar learning.

## 1.2. Cognitive theories of Artificial Grammar Learning

Several studies have demonstrated that humans can perform above chance in Artificial Grammar Learning paradigms. However, it is still debated what determines participants' behavior, and different theoretical accounts have been proposed (Pothos, 2007). The first theoretical description was provided by A. Reber (Reber, 1967), who suggested that participants learn the grammar' rules implicitly. Successive studies have questioned this interpretation, on the grounds of experimental evidence pointing at several potential confounds (Brooks & Vokey, 1991; Shanks & Stjohn, 1994; Vokey & Brooks, 1992). A revised account of the same hypothesis suggested that participants do not learn the full spectrum of grammatically correct transitions, but only a subset of it, i.e. microrules, such as bigrams or trigrams (Dulany, Carlson, & Dewey, 1984; Reber & Lewis, 1977). Similarly, the chunking hypothesis suggests that the more frequent chunks are the more salient, and consequently better learnt by participants as grammatically correct (Knowlton & Squire, 1996; E. Servan-Schreiber & Anderson, 1990). A further computational perspective posits that humans learn sequences as recurrent models, that is, by decoding relevant features from former adjacent items in the string (Boucher & Dienes, 2003; Cleeremans & McClelland, 1991; Cleeremans, Servan-Schreiber, & McClelland, 1989). Overall, contrary to human experiments in which subjects typically see a few dozen examples, all considered computational models have been trained with large datasets, and sometimes with significant overlap between training and test sets, making it difficult to draw any substantial comparisons with human cognition.

## 1.3. Implicit and explicit learning

Another important aspect investigated in the experimental AGL literature regards the distinction between implicit and explicit learning. There are aspects of knowledge for which we have little conscious introspection, such as language (Reber, 2008). Experimental evidence has suggested that AGL occurs implicitly, that is in an automatic and nonintentional way (Cleeremans, Destrebecqz, & Boyer, 1998; Reber, 1967; Squire & Dede, 2015). Despite skepticism about the very existence of implicit learning (Newell & Shanks, 2014; Shanks & Stjohn, 1994 but see in opposition Alamia et al., 2016), the existence of distinct implicit and explicit systems in the human brain has been postulated, the former being automatic, the latter being flexible and directed to hypothesis testing (Cleeremans, 2005; Koch, 2004). In addition, previous AGL experiments demonstrated that more complex grammars are more likely to be processed implicitly than simpler grammars (Halford, Wilson, & Phillips, 1998; Reber, 1976, 1989). This result is in accordance with the assumption that working memory has restrictions on the number of items (i.e. hypotheses) explicitly accessible at any given time (Capacity, 2007; Cowan, 2010), thus limiting abilities to process complicated rules (e.g. based on long-distance dependencies). On the contrary, implicit processes possibly rely on different mechanisms, capable of dealing with larger amount of information at the cost of reduced flexibility (Rah, Reber, & Hsiao, 2000; Röttger, Haider, Zhao, & Gaschler, 2017; Smith & McDowall, 2006).

### 1.4. Purpose of the study

In this study we tested 4 grammars spanning over 3 Chomsky's hierarchy levels. Both human participants and artificial neural networks were trained and tested on datasets generated from those grammars. Importantly, we aimed to use comparable amounts of training for humans and artificial neural networks. Our purpose was to investigate which architecture—feed-forward vs. recurrent networks—better captures human behavior as a function of grammar complexity. Moreover, as AGL is an established framework to contrast implicit and explicit learning (Dienes & Perner, 1999; Dienes & Berry, 1997), we aimed at testing whether these modes could be related respectively to feedforward and recurrent architectures, similarly to findings in visual perception (Lamme & Roelfsema, 2000; Supèr, Spekreijse, & Lamme, 2001).

## 2. Materials and methods

### 2.1. Artificial grammar datasets

We performed 4 experiments with different artificial grammars (fig.1A), each composed of the same amount of correct and incorrect sequences. According to the Chomsky hierarchy introduced above, two grammars were *regular*, referred in the following as grammar A and grammar B, one was *context-free* and one was *context-specific*. In the **regular** grammars, the sequences' length ranged from 2 (grammar A) or 3 (grammar B) to an arbitrary maximum length of 12 items. Longer sequences were discarded. Grammar A's vocabulary was composed of 4 letters, while grammar B counted 5 letters. The dataset from **context-free** and **context-specific** grammars (respectively type II and type I of the Chomsky hierarchy) were composed following an approach based on symmetry (Öttl et al., 2015; Tecumseh Fitch & Friederici, 2012). In both cases the vocabulary was composed of 10 letters combined in 5 pairs, and the strings' length was either 4, 6 or 8. Figure 1A shows in details how sequences were generated in each grammar.

### 2.2. Humans

#### 2.2.1. Participants

Overall, 56 participants (31 female, age=25.4 ± 4.7) took part in 4 experiments using 4 different artificial grammars (n=15 for Grammar A, B and Context-Free; n=11 for the Context-Specific grammar). All participants gave written consent before the experiment, in accordance with the Declaration of Helsinki and received monetary compensation. This study was carried out in accordance with the guidelines for research at the "Centre de Recherche Cerveau et Cognition" and the protocol was approved by the committee "Comité de protection des Personnes Sud Méditerranée 1" (ethics approval number N° 2016-A01937-44).

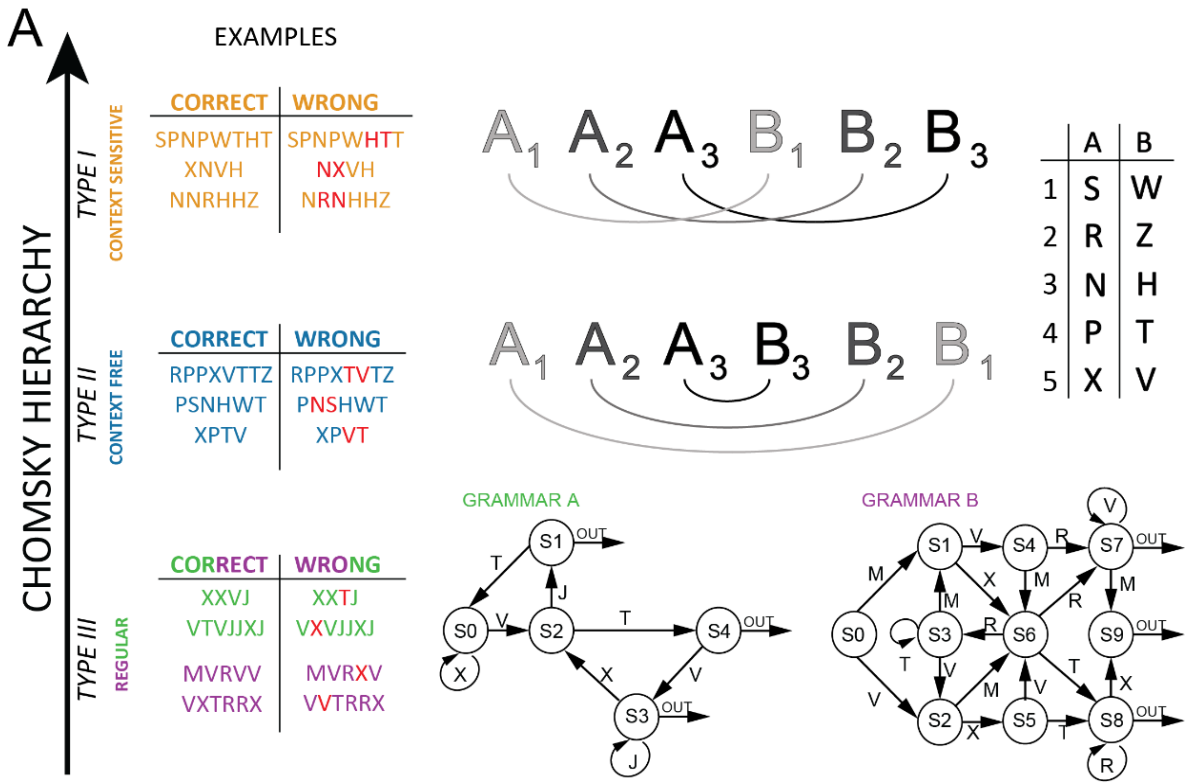

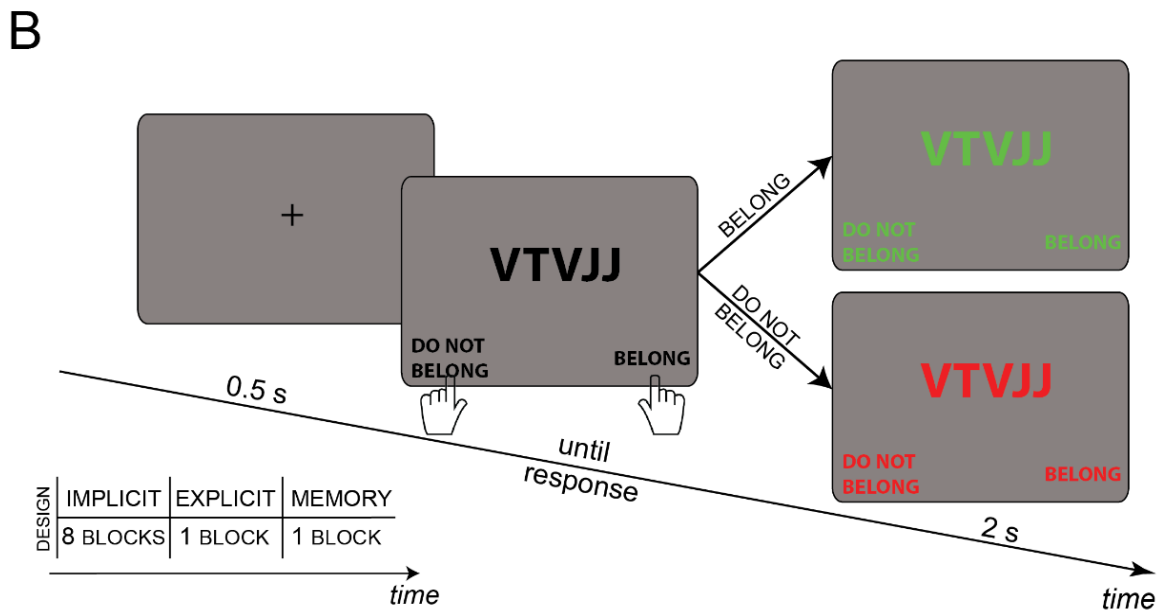

**Figure 1**. A) A schematic representation of the 4 grammars employed in this study, arranged according to the Chomsky hierarchy. **Regular** grammars are defined by a scheme organized in directionally connected nodes (type III, the lowest in green and purple). For both grammars A and B, each correct sequence was generated starting at node S0 of each respective scheme, and then randomly choosing an arc departing from that node and recording its letter. This procedure was iterated until one of the arc labeled as 'out' was reached, thus terminating the string. Incorrect sequences had only one inexact item, being substituted by another letter from the proper vocabulary but violating grammatical rules (in red in the table). In type II or **Context-Free** grammar (CF, in blue), the second half of the sequence mirrored the first half, applying the respective pairing to each letter (e.g. if A3 was the letter N, then the letter B3 was H). Vice versa, in type I or **Context-Specific** grammar (CS, in orange), a translation was applied from the first to the corresponding second half of the string. For example, considering the first half as 'PPN' and referring to the same pairing as in the picture, the corresponding mirrored version would be 'HTT' in the CF, and 'TTH' in the CS grammar. The incorrect sequences were obtained by switching only two non-identical letters within one of the two halves (in red in the table). Note that a correct string in the context free grammar is considered as incorrect in the context specific, and vice versa. B) Time-course of each trial. All the human experiments employed the same design. The sub-table reports the number of blocks for each session: implicit, explicit and memory. See section 2.2.2 "Experimental design" for details.

### 2.2.2. Experimental design

The same experimental design was applied for each grammar. Each trial started with a fixation cross lasting 500ms, followed by a string of letters displayed in the center of the screen (fig.1B). Participants were informed that there were two groups of respectively correct and wrong sequences. They were asked to classify each sequence by pressing one of two key-buttons (respectively with the right and left index). Participants were not explicitly instructed about the existence or the nature of the rules generating the sequences. No time constraints were imposed to provide an answer. As soon as the response was given, visual feedback was provided: the string turned green when correctly classified, red otherwise. The visual feedback lasted for 2 second before starting the next trial. At each trial we recorded accuracy and reaction times.

Each participant performed one session lasting approximately 1-hour and composed of 10 blocks. During the first 8 blocks, labeled as *implicit* in the sub-table in figure 1B, participants were not explicitly informed about the existence of the rules generating the sequences. Each block of the *implicit* part counted 60 trials, for 480 trials in total. A questionnaire was provided between the 8$^{th}$ and the 9$^{th}$ block to assess participants' explicit knowledge of the rules. The questionnaire was different for each grammar, asking specific questions about the rules (see appendix C). In grammar A and B participants responded to 7 multiple-choice questions, whereas in CF and CS grammars participants were asked to point to the wrong letter in a series of 7 novel sequences. In all grammars, participants were asked to report their confidence level from 0 to 100 after each answer. Following the questionnaires participants were asked to report (part of) the rules at the best of their knowledge. The last 2 blocks (labeled respectively as *explicit* and *memory* in figure 1B) were identical to the previous ones, but served as control conditions. In the 9$^{th}$ block, composed of only 20 trials, participants were provided with a printed scheme explaining exactly the rules of the grammar (the same as reported in figure 1A), and were instructed about the generation of correct sequences. During this block, they were allowed to consult the scheme before providing each answer. In the 10$^{th}$ and last block, participants were asked to perform the same task for additional 20 trials but no longer had access to the grammar scheme, thus supposedly relying on their memory of the rules.

### 2.3. Artificial Neural networks
#### 2.3.1. Experimental design

The neural network design was composed of two parts: a first parameter search, and a subsequent comparison with human behavior. Both were implemented using the Keras library (Chollet, 2015), back-ended in Tensorflow (GoogleResearch, 2015). Altogether, we trained feedforward and recurrent architectures, each composed of a series of fully connected layers. All networks were trained to classify the sequences as correct or wrong, employing the same dataset (i.e. 4 grammars) and the same amount of trials as in the human experiments.

Regarding the parameter search, we aimed at determining the parameters whereby each architecture scored closest to human performance. We tested a range of networks, varying the number of layers and the learning rate, defining a 2-dimensional space (see fig.3). The training set was composed of 500 sequences (roughly similar to humans, who viewed 480 training examples), whereas the validation and testing set were composed respectively of 100 and 200 sequences. All the layers of a given network counted the same number of neurons except the output layer, which had only one neuron. The number of neurons was chosen such that all networks within the same 2-dimensional space had (roughly) the same number of free parameters. We explored 4 possible spaces with different numbers of parameters: respectively 1400, 7900, 31,000 and 122,000. One axis of the space referred to the number of layers, counting 6 levels (i.e. 2, 3, 4, 5, 7 and 10), whereas the other axis represented the learning rate, which counted 20 levels. Different values were used for the two families of architectures (see below). Each parameters space counted 6x20=120 networks, each one trained 20 times with random weights initialization. At first, we determined which networks provided the closest-to-human performance. For each grammar, we averaged between subjects the performance on the last block, and we subtracted this value from the mean performance of each network computed over 200-sequences (test set) after a learning of 500 examples (training set). For each parameters-space and grammar, we selected the network with the smallest absolute difference as the one closest to human behavior. Note that the selected network is not necessarily the one with the highest performance on the test-set (see fig.A1 in appendix A). Once we determined the closest-to-human networks, we obtained their respective learning curves by varying the training set size progressively from 100 to 500 sequences, with a stride of 100. As in the parameter search, we averaged the results over 20 random weights initialization, using respectively 100 and 200 sequences for the validation and test set.

### 2.3.2. Feedforward architectures

Feedforward neural networks were composed of fully connected dense layers. The input layer counted 12xK neurons, representing the one-hot encoding of the 12-letters longest possible string (K represents the total number of letters, equal to 4, 5 and 10 for grammar A, B and CFG/CSG respectively). We employed zero-right padding when shorter sequences were fed to the network. All activation functions were defined as rectified linear units, i.e. 'ReLu', except the output neuron, which was implemented with a sigmoid function. The loss function was defined as 'mean-square-error', and optimized by means of stochastic gradient descent (Robbins & Monro, 1985) with Nesterov momentum (Nesterov, 1983) set to 0.9, and decay equals to 1e-06. In all grammars, both in the parameter search and in the learning curve estimation, we considered 1 epoch only (500 trials), with batch size of 15.

### 2.3.3. Recurrent architectures

Recurrent neural networks were composed of fully recurrent connected layers, in which each neuron was connected to itself and all other neurons in its layer. Starting from each sequence's first letter, at each time step the following letter was provided to the network as a one-hot encoded vector. The input layer thus counted as many neurons as letters in the grammars alphabet. A sigmoid activation function armed the output neuron, whose activation determined the classification decision after the last letter of the string was fed to the network. Learning occurred via back-propagation through time (Mozer, 1989; Werbos, 1990). As in the feedforward architecture, we considered 'mean-square-error' as loss function, optimized with the keras function rms-prop (Tieleman, Hinton, Srivastava, & Swersky, 2012). We set rho to 0.9, epsilon to 1e-8 and decay to 0. As for the feedforward networks, we employed only 1 epoch (500 samples) and batch size of 15 in both the parameter search and the learning dynamic part.

## 2.4. Data analysis

In the human experiments, we recorded accuracy and reaction times at each trial. Both measures were averaged over blocks and tested by means of Bayesian ANOVA (JASP Team, 2018; Love et al., 2015). Each analysis provides a Bayes Factor (BF) for each independent factor (e.g. BLOCK). As widely accepted (Kass & Raftery, 1995), we considered BF above 3 as substantial evidence in favor of the alternative hypothesis, over 20 as strong evidence, and BFs beyond 100 correspond to very strong evidence. Conversely, a BF below 0.3 suggests lack of effects (Bernardo & Smith, 2008; Masson, 2011). In all our analyses we reported BFs and estimate's error.

# 3. Results

## 3.1. Human results

### 3.1.1. Accuracy and Reaction Times

For each grammar we assessed whether participants learned the rules by testing accuracy and reaction times (RT) by means of a Bayesian ANOVA, considering BLOCK as independent factor (categorical, from 1 to 8) and SUBJECT as a random factor. At every level of the Chomsky's hierarchy, participants learned the rules above chance (all BF $\gg$ 100 for the BLOCK factor, for each grammar), as shown in figure 2A. Conversely, reaction times did not show any significant effect at any level of the Chomsky hierarchy, remaining between 2 and 3 seconds during the whole experiment. We found very similar results in both accuracy and RT in analyzing the sub-set of participants who failed in reporting the rules during the questionnaire in CF and CS grammars (respectively N=8 and N=12, in darker colors in the figure). Regarding grammars A and B, all the participants failed in reporting any subset of the rules.

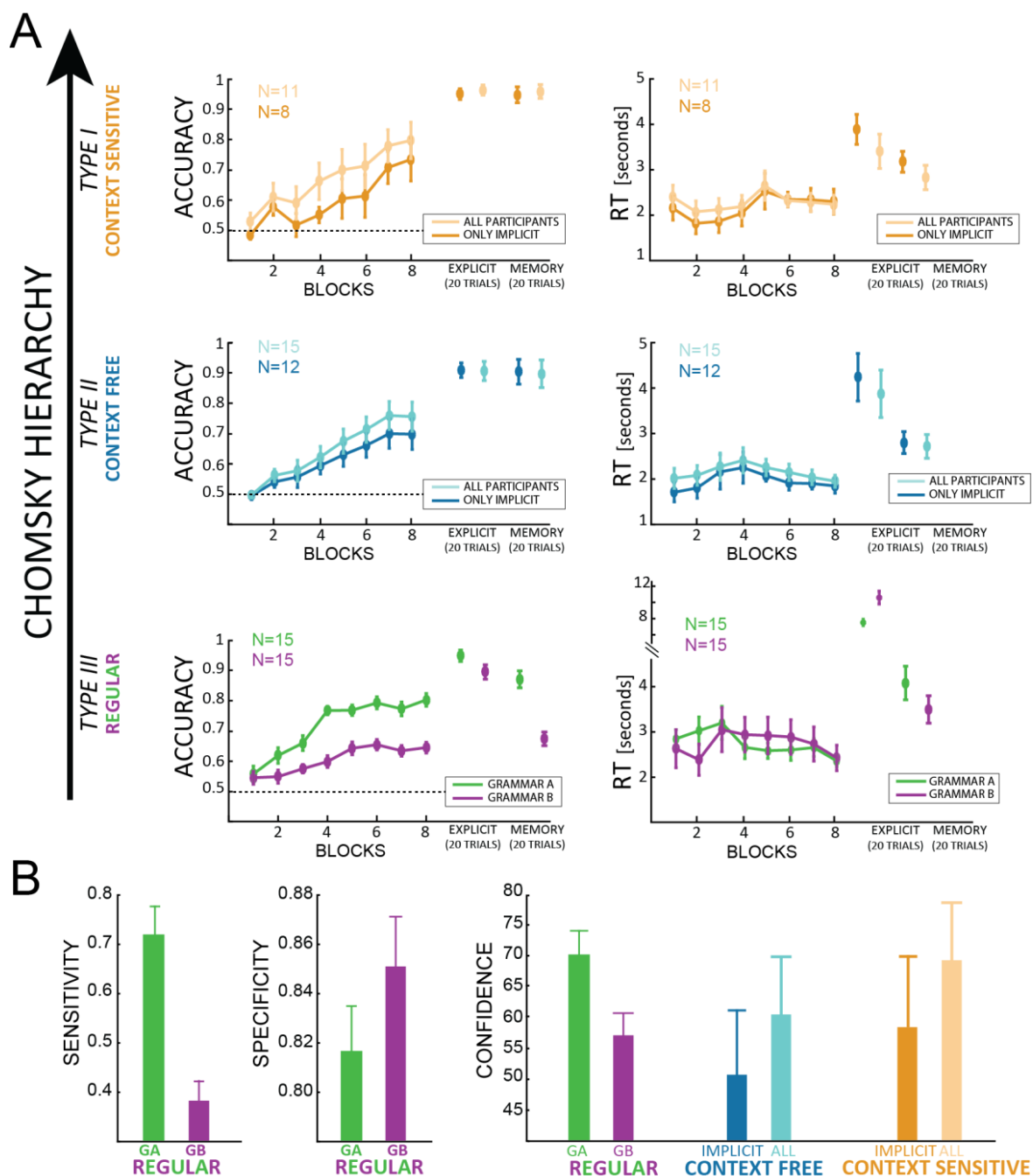

**Figure 2**. A) For each level of the Chomsky hierarchy, accuracy (left column) and reaction times (right column) are plotted as a function of the implicit BLOCKS, followed by the EXPLICIT and MEMORY blocks. In type I and type II (first 2 rows), darker colors represent the sub-set of "implicit" participants (i.e. those who failed to report the rules during the questionnaire), lighter colors all the participants. In type III, green and purple represent respectively grammar A and B. All values are expressed as average over participants ± standard errors (N values reported in each subplot). B) Sensitivity and specificity indexes for explicit report of the grammar rules are expressed as average ± standard errors for grammar A (green) and B (purple). The right subplot shows the average confidence values expressed by all participants (darker colors in CFG and CSG represent the implicit sub-set of participants).

Furthermore, we performed a Bayesian ANOVA to contrast the *explicit* and *memory* blocks with the 8th block of the *implicit* session (categorical factor CONDITION, 3 levels), considering only the implicit sub-set of subjects in CFG and CSG, and all the participants in grammars A and B. This analysis revealed a significant difference between conditions at every level of the Chomsky's hierarchy (all BF>>10) with larger differences between *implicit* and *explicit* blocks. Moreover, the accuracy was similar between the memory and the explicit tasks in all level of the Chomsky hierarchy (all BF<<3) except grammar B (BF=7272, error=9e-9%), suggesting that once participants knew the rules they could recall them shortly afterwards (full results table in appendix B, table B.1). All in all, the difference between *implicit* and *explicit* blocks suggests that participants were not fully aware of the rules during the *implicit* blocks, as their behavior changed when they could inspect the grammars' scheme. This conclusion is further corroborated by the subsequent questionnaires' analysis.

### 3.1.2. Questionnaire

Participants filled a questionnaire after the $8^{th}$ block of the experiment (end of the *implicit* part). Regarding the *regular* grammars A and B, each question conceded 5 answers, having more than one possibly correct (see appendix C1). For each question we computed the Sensitivity and the Specificity, defined as the proportion of positive (Sensitivity) or negative (Specificity) responses correctly identified as such (Cadogan, 2012). Interestingly, we observed a significant difference between grammar A and B in the Sensitivity index (BF>100, error=7.69e-7), revealing a more explicit knowledge in grammar A than B (figure 2.B). No difference was observed in the Specificity index (BF=0.647, error=0.003). Regarding CF and CS grammars, participants had to first indicate the wrong letters in 7 incorrect sequences, and then report the rules or pattern they followed to perform the task (see appendix C.2). Participants who correctly reported the rules' scheme as represented in figure 1A (3 participants for both grammar CFG and CSG) were also able to detect the wrong letters in the 7 incorrect sequences provided in the questionnaire. Contrarily, participants who reported no rules or wrong ones, were unable to detect the wrong letters in the incorrect sequences (they provided either wrong or no answers). In CF and CS grammars we considered as 'implicit' the participants who failed in reporting the rules at the end of the experiment. Lastly, in all grammars we compared the averaged confidence report provided after each question (figure 2.B). Overall, we did not observe any significant difference between the grammars (BF=0.216, error=4.6e-4; very similar results were obtained considering the subgroup of "implicit" participants –see above- in grammars type II and I).

### 3.2. Artificial Neural networks results

Regarding the neural networks analysis, we first performed a parameter search in order to identify the networks whose performance was closest to the human one. We tested 120 networks for each parameter space (each one defined by 2 axes: learning rate and number of layers – see methods sub-section 2.3.1). Each network was trained with a similar number of trials as in the human experiment (i.e. 500 compared to 480 in humans), and then tested over 200 sequences (with frozen parameters, i.e. no training). Figure 3A shows the absolute difference between human performance in the last block (i.e. trials from 420 to 480) and the networks performance after a training period of 500 sequences. Only the results regarding the 31.000 parameters space are shown, but similar results were obtained considering all other spaces (see in appendix A figure A.1). Notice that the networks closest to human performance (i.e. whose absolute difference is the smallest) are not necessarily the one performing the best. Specifically, regarding the feedforward networks, we observed that the best results were obtained with the lowest number of layers (corresponding to the highest number of neurons per layer), at each level of the Chomsky's hierarchy. Since the feedforward networks underperformed when compared to human performance, the ones with the best performance were also the closest to human's behavior. Concerning the recurrent networks, we observed that the higher accuracies corresponded to lowest learning rates, and that in regular grammars, the closest-to-human network did not correspond to the one with the best performance (the difference in performance between the best and closest-to-human networks are 0.02 and 0.08 for grammars A and B respectively). Additionally, we averaged the results from all the levels of the Chomsky hierarchy to determine which neural network was the closest to human behavior irrespective of the rules' complexity level (fig.3 C,D – see appendix A figure A.2 for performance over trials). The results confirm that feedforward architectures perform best with the lowest number of layers, whereas recurrent architectures achieve the best accuracy with lower learning rate. Finally, we obtained the learning curves for the 8 best-performing models (4 grammars * 2 network types) in order to compare them with the human ones.

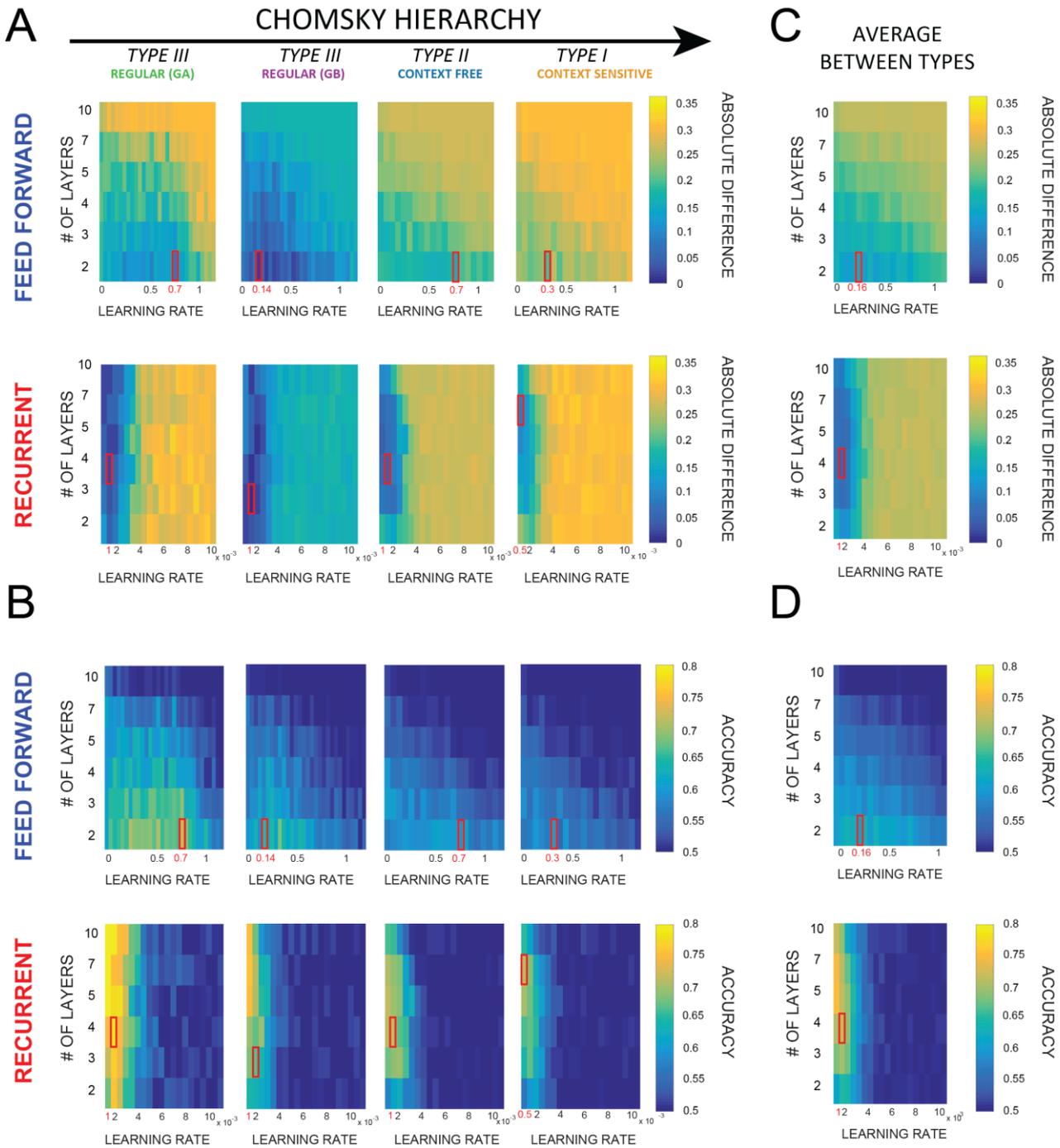

**Figure 3**. Results of the parameter search for the third space (i.e. 31,000 parameters). Each 2D space is defined by two axes: learning rate (LR) and number of layers (NL). Each point in the space represents the averaged performance of the network with specific LR and NL. The red square highlights the network closest to human performance for each plane, as the ones with the smallest difference (between NN and human performance) at the end of training, as shown in panel A. In all subplots, upper rows are feedforward networks, lower rows are recurrent ones. A-B) Each column represents one grammar, ordered according to the Chomsky's hierarchy. Subplot A shows the difference between NN and human performance, i.e. the distance to human performance at the end of training (see sections 2.3.1 for details), subplot B the accuracy of each network. Similar results are observed between grammars: in FF networks 2 layers are sufficient to obtain the best results, whereas recurrent networks perform best with the lowest LR. C-D) Same as A-B but considering the average across all grammars.

### 3.3. Human-networks comparison

To investigate whether not only the final accuracy, but also the temporal dynamics of training are comparable between humans and artificial neural networks, each network selected based on its final performance after 500 training trials (Figure 3) was re-trained with 5 different training-set sizes, ranging from 100 to 500 trials, and each time its performance was tested over 200 novel strings. The results are shown in figure 4 and represent the average ± standard error over 20 repetitions. For the human learning curves, we averaged a 40-trials long window centered over the trials corresponding to the same size as the training set. The first/last bins were computed averaging the first/last 40 trials of the first/last block. We performed a Bayesian ANOVA considering as categorical factors TRIAL (from 1 to 5, corresponding to the size of the training set), AGENT (three values representing humans, feedforward –FF- and recurrent –RR- architectures) and GRAMMAR (4 levels corresponding to the 4 grammars we tested). The results suggested a robust effect of each factor (all BF>>3e+15, error<0.01%), and a very strong interaction between the factors AGENT and GRAMMAR (BF=2.3e+14, error<0.01%). In order to disentangle this significant interaction, we tested a post-hoc analysis for each grammar separately, considering as factors only TRIAL and AGENT. Interestingly, in the highest level of the Chomsky hierarchy (type I and type II) we observed a significant difference between humans and FF (all BF>1000, error<1e-6%) and between RR and FF (all BF>1000, error<1e-6%)), but not between humans and RR (all BF<1, error<0.001). In type III grammars we reported two different patterns of results: in grammar A we observed results similar to the previous ones, having a significant difference between all the agents (human-FF and RR-FF: BF>1000, error<1e-6%; human-RR, BF=10.53, error=2.5e-7). However, such a difference did not emerge in grammar B, in which we observed no difference between AGENTs (BF=0.79, error=0.02%). In summary, as shown in figure 4C, the pattern of results suggests that recurrent architectures are closer to human behavior at every level of the Chomsky hierarchy, with the exception of grammar B, for which recurrent and feed-forward models cannot be distinguished, as further discussed in section 3.4.

Moreover, we also compared models and humans performance as a function of sequence length (fig.4 B,D). For each grammar, we tested a Bayesian ANOVA having as factors sequence LENGTH, and AGENT (human, FF or RR). Interestingly, we found a strong effect of the factor LENGTH for each grammar (all BF>>1000, error<0.4%), revealing that human participants and both model architectures performed better with shorter sequences (table B.2 in appendix B for full results). Moreover, in all except grammar B we found a very significant difference between AGENTs (grammar B, BF=0.38, error=0.022%; all other grammars BF>1000, error<0.9%). Post-hoc analyses confirmed our previous results, revealing that recurrent architectures matched more closely human data than the feedforward ones.

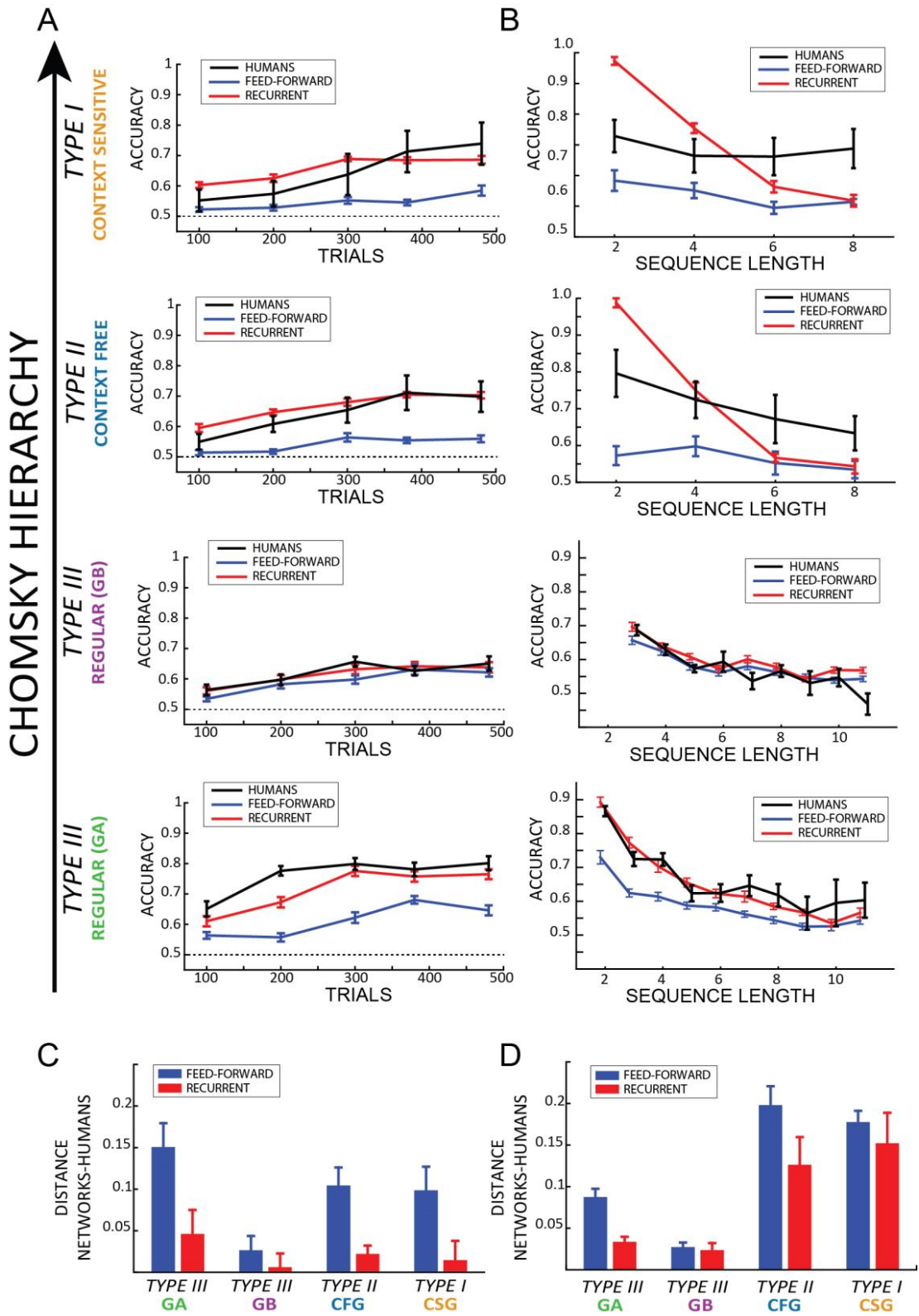

**Figure 4**. A,B) Results over trials (A) and string lengths (B) for humans (in black) feedforward (in blue) and recurrent (in red) networks. For humans in (A) each bin is an average over 40 trials (20 trials before and after respectively, except the last bin which includes the last 40 trials of the experiment). Each row represents a grammar, ordered according to the Chomsky's hierarchy. C-D) The plots show the distance between humans' performance and FF (in blue) and RR (in red). Each distance represents the area measured between the human and the network curves. Except for grammar B, RR networks are significantly closer to human performance, both as a function of training time and sequence length.

### 3.4. Difference between feedforward and recurrent networks in regular grammars

Previous results suggested that recurrent architectures perform better in modelling human learning than feedforward ones at every level of the Chomsky hierarchy. However, one of the regular grammars we tested (i.e. grammar B) violates this conclusion, revealing no significant differences between the two architectures and the human behavior. In order to shed some light on this result, we collected 8 additional artificial regular grammars from a recent review (Katan & Schiff, 2014) and tested both our recurrent and feedforward architectures on each grammar. Based on our data, in which a simpler grammar (GA) leads to a larger difference between FF and RR networks than a more complicated one (GB), we hypothesized that RR networks would perform better (i.e., closer to humans) than FF networks in simpler grammars. Consequently, we defined 5 simple metrics to characterize the complexity of each grammar: number of letters in the vocabulary; number of states; number of transitions (or rules); number of bigrams and shortest length of a sequence (or minimum length).

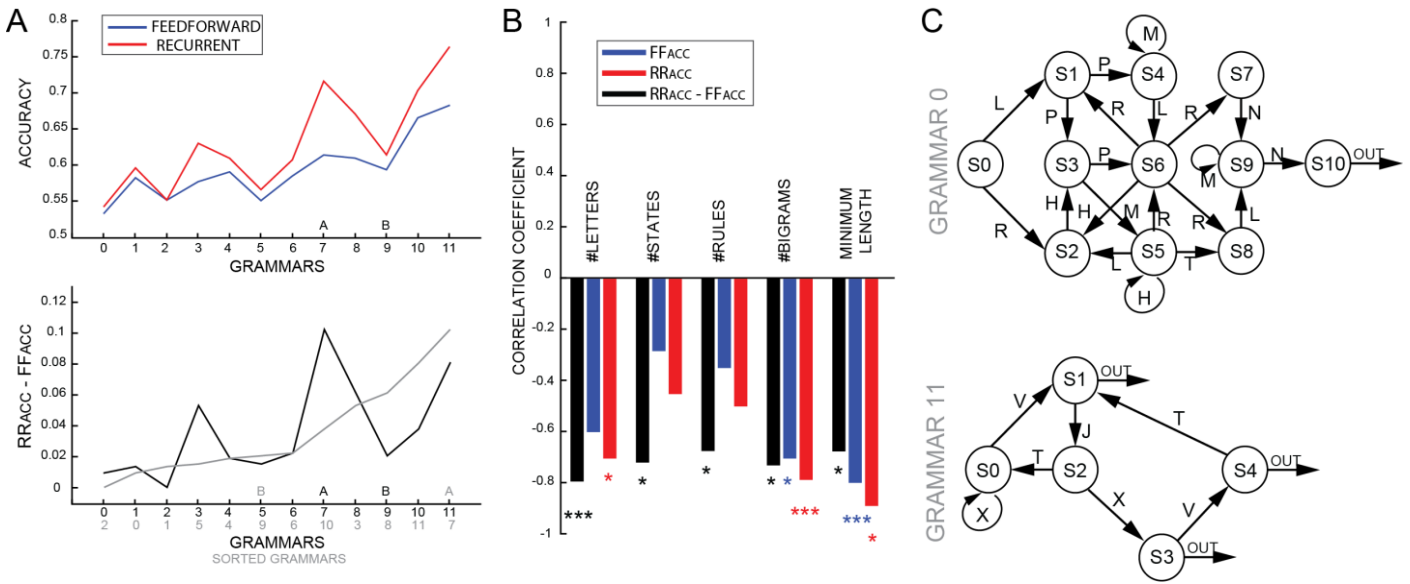

**Figure 5**. A) The upper panel shows the performance of FF (in blue) and RR (in red) architectures for each grammar (from 1 to 10 of the x-axis), averaging over 20 initialization over a training set of 500 trials. Grammars A and B are respectively 7 and 9. The difference in the performance between the two architectures is shown in the lower panel (the same sorted difference in light grey). B) Pearson indexes obtained correlating the performance of FF and RR networks and their difference (respectively in blue, red and black) with the 5 complexity metrics across the 10 grammars. C) The two novel grammars we designed at the 2 extremes of the complexity index to test our hypothesis.

Overall, the recurrent network always performed better than the feedforward one for all 10 grammars (fig.5A, upper panel). However, the difference between the two architectures was not constant across grammars. As shown in figure 5B, such difference correlated significantly (and negatively) with each of the complexity metrics (all BF>5, error<0.01%), suggesting that the difference between the two architectures is inversely correlated with the complexity of the grammars (see in appendices figure A.3 and table B.3 for the correlation matrix). Recurrent networks outperform feedforward ones in simpler, but less so in more complicated grammars.

To confirm this conclusion, we created two novel original grammars (fig.5C), one at each extreme of the complexity metrics we defined above, and we trained feed-forward and recurrent models on these new grammars. As predicted, the difference between FF and RR was large (0.11%) in the simplest and small (0.01%) in the more difficult grammars (numbered respectively 11 and 0 in fig.5A).

## 4. Discussion
### 4.1. Summary of the results

In this study we demonstrated that recurrent neural networks mimic human artificial grammar learning more closely than feedforward architectures, irrespective of the grammar's level within the Chomsky hierarchy. This result supports the hypothesis that recursion is key in cognitive processes such as language (Fischmeister, Martins, Beisteiner, & Fitch, 2017; Jackendoff, 2009, 2011; Steedman, 2002). As already mentioned, previous studies showed that humans can learn to classify sequences as correct or not according to grammatical rules, and such knowledge appears to be to some extent implicit (Knowlton & Squire, 1996; Reber, 1967; Rohrmeier, Fu, & Dienes, 2012). However, in a usual AGL experiment, participants are first asked to memorize a set of sequences (training phase) and then to classify a new set as correct or not (testing phase). Here, we combined the two phases such that training and testing occur at the same time, allowing us to track the learning dynamics as it progresses. This design let us compare the participants' learning with the artificial networks' one. Importantly, we showed that both feedforward and recurrent neural networks can learn artificial grammars within the same limited number of trials as for human participants. However, the overall behavior of the recurrent networks, and in particular their learning dynamics, was closer to human behavior.

### 4.2. Related work on artificial grammars and neural networks

In our simulations, we contrasted the performance of feedforward and recurrent architectures in AGL tasks. The two architectures are substantially different, and represent distinct functional processes. Feedforward networks allow the information to flow one way from input to output. In feedforward architectures the sequence is processed at once, with a single sweep. A very similar approach is the competitive chunking network (Servan-Schreiber & Anderson, 1990), in which the whole sequence is processed as one to identify hierarchical, nested patterns (i.e. the chunks). In this model several layers can be involved in the process before classification occurs (Buchner, 1994). However, our results suggest that FF networks having only 2 layers provide the best classification in AGL tasks within our experimental constraints (with a limited training dataset, comparable to human learning). Possibly, deeper networks would improve classification performance given a larger amount of trials, and may be more powerful with more complex grammars; however, we lack experimental studies showing how humans would learn these more complicated structures. On the other hand, recurrent models process the sequence differently, receiving one letter at a time and building short and long temporal dependencies between items (Elman, 1990). The learning occurs by reinforcing associations that appear frequently within the training set, and it has been applied successfully to finite state grammars since the early '90s (Cleeremans et al., 1989; Dienes, 1992; Giles et al., 1992). Since these early studies investigating connectionist models of artificial grammar learning, recurrent neural networks have been proposed as an ideal candidate to investigate how learning occurs in artificial grammar experiments, particularly in the context of implicit learning ( Cleeremans et al., 1989; Servan-Schreiber, Cleeremans, & McClelland, 1991). In particular, two properties have been considered compelling in recurrent neural architectures: first, their sensitivity to temporal structure; second, the fact that the rules' representations are distributed over the weights of the networks, as assumed in implicit learning models (Cleeremans, 2007; Cleeremans & McClelland, 1991). Previous studies have investigated how neural networks learn to encode the states of Finite State Automata (FSA, equivalent to regular grammars in Chomsky's hierarchy), demonstrating that after learning each automaton (i.e. grammar) state is represented in the activation of the recurrent layers (Cleeremans et al., 1989). Subsequent work expanded this conclusion, demonstrating that second-order recurrent networks can infer the full FSA after an appropriate amount of training examples or epochs (Giles et al., 1992). Context-free and context specific grammars have also been studied by means of simple recurrent networks. One study proved that SRNs can learn to predict the next letter in a simple CF language, in which sequences are composed of two groups of letters ('A' and 'B') and organized according to the structure $A^nB^n$. The authors, interpreting recurrent networks as dynamical systems (Petersson, Grenholm, & Forkstam, 2005), demonstrated that the model learns to count items from each group in order to successfully determine the grammar's rules (Rodriguez, Wiles, & Elman, 1999). In a later study, they extended the results by showing that recurrent networks process and store information as if using a memory buffer, thereby solving both CF and CS grammars (Rodriguez, 2001). Another study investigated how Long Short Term Memory (LSTM) recurrent networks learn CF and CS grammars (Gers & Schmidhuber, 2001). These LSTM architectures are becoming more prominent in natural language studies, given their ability to learn long-distance dependencies. However, in this and all previous studies, neural networks were trained on a very large number of examples over several epochs, as summarized in table 1. This approach prevents a fair comparison between models and human performance, as human participants are able to classify novel strings above chance after only a few hundred examples.

| AUTHORS | GRAMMAR | NETWORK | TOTAL TRAINING EXAMPLES |
|---|---|---|---|
| Cleeremans et al. (1989) | Regular | Recurrent | 60,000 |
| Servan-Schreiber, Cleeremans, & McClelland (1991) | Regular | Recurrent | 60,000 |
| Cleeremans & McClelland (1991) | Regular | Recurrent | 60,000 |
| Giles et al. (1992) | Regular | Recurrent | 150,000 |
| Dienes (1992) | Regular | Recurrent | 20,000 |
| Rodriguez, Wiles, & Elman (1999) | CFG | Recurrent | ~15,000 |
| Bodén, Wiles (2000) | CFG / CSG | Recurrent | 10,000 |
| Kinder, Shanks (2001) | Regular | Recurrent | 2880 |
| Gers, Schmidhuber (2001) | CFG / CSG | Recurrent (LSTM) | 10e3 to 10e7 |
| Boucher, Dienes (2003) | Regular | Recurrent&Chunking Model | ~1000 |
| Tunney, Shanks (2003) | Regular | Recurrent | 3200 |
| Petersson, Grenholm, & Forkstam (2005) | Regular | Recurrent | 25,400 |
| Wierzchon, Barbasz (2008) | Regular | Feedforward | 96,000 |
| Kinder, Lotz (2009) | Regular | Recurrent&Chunking Model | ~1000 |
| Duarte, Seriès, & Morrison (2014) | Regular | Recurrent (spiking) | 10e3 to 10e5 |
| Cohen, Caciularu, Rejwan & Berant, (2017) | Regular | Recurrent | 225,000 |
| Alamia, Gauducheau, Paisios, VanRullen (2019) | Regular / CFG / CSG | Feedforward & Recurrent | 500 |

**Table 1**. Representative overview of studies training artificial neural networks to classify novel strings. Most studies employed recurrent networks over regular grammars. The last column provides the total number of sequences the network was trained on (sequences per epoch, times the number of epochs). Our study, highlighted in yellow, employs the smallest number of training examples, a few orders of magnitude smaller than most studies, and similar to typical experimental studies in humans.

### 4.3. A fair comparison between human and artificial neural networks.

Table 1 provides a representative overview of studies investigating the behavior of artificial neural networks in classifying sequences from different grammars. First, it appears that previous studies have investigated mostly recurrent rather than feedforward networks, even though implicit learning has been frequently related to feedforward processes (Boly et al., 2011; Koch, Massimini, Boly, & Tononi, 2016; Lamme & Roelfsema, 2000; see also next section 4.4). Second, all but two studies employed thousands (or tens of thousands) of sequences to train the models' parameters (except Boucher & Dienes, 2003 and Kinder & Lotz, 2009, which used ~1000 examples across the training epochs). In our study, highlighted in yellow, we implemented a fairer comparison with human performance, as both artificial networks and human participants were trained on the same limited number of examples (~500). Moreover, differently than previous studies on artificial grammar learning in humans, we adopted an experimental design in which training and testing occurred at the same time, allowing a direct comparison of learning dynamics (i.e., learning curves) between humans and artificial neural networks. Previous studies had compared recurrent networks and a chunking model (similar to a feedforward architecture, see paragraph 4.2) with human behavior, but with opposite results (Boucher & Dienes, 2003; Kinder & Lotz, 2009). Our approach allows us to directly compare human performance with both feedforward and recurrent models, avoiding the limitations discussed above. All in all, this comparison reveals that recurrent models perform closer to humans than feedforward ones, except in more complicated –and supposedly implicit- grammars. For those grammars (e.g. Grammar B), human performance remains poor, and can equally well be accounted for by recurrent or feedforward models. Moreover, another consideration that emerges from our comparison regards the crucial role of the learning rate in recurrent networks. Interestingly, in an attempt to model the behavior of Amnesic and Healthy subjects during an artificial grammar learning task, some authors reported a steady decrease in the classification performance of a simple recurrent network with the increase of its learning rate (Kinder & Shanks, 2001)— as if Amnesic patients had different learning rates compared to Healthy subjects. Their conclusions are in line with our observations, as they confirm that recurrent networks perform best in classifying novel sequences with low learning rate values. Finally, another appealing result of our investigation concerns the functional distinction between implicit and explicit processes in artificial grammar learning, as examined in the next section.

### 4.4. Computational correlates of implicit and explicit processes

In the two regular grammars we observed a significant difference in participants' awareness of the rules. In grammar A participants performed better at the questionnaire than in grammar B, coherently with the hypothesis that simpler grammars are more likely to be learnt consciously (Halford et al., 1998; Reber, 1976; Sun & Peterson, 1994). The distinction between implicit and explicit processes has been expressly designed in an integrated model tested also on artificial grammar tasks (Sun, 2006; Sun, Zhang, Slusarz, & Mathews, 2007), providing evidence in favor of the hypothesis that implicit processes precede explicit ones (Cleeremans, 1997; Windey & Cleeremans, 2015) and are prominently involved in complex grammars (Halford et al., 1998; Reber, 1976). Interestingly, our results reveal that in the more implicit grammar A, both ANN architectures reliably tracked human behavior, whereas only recurrent networks achieved this goal in more explicit grammars. This result draws a compelling parallel between feedforward/recurrent and implicit/explicit processes, consistently with results in visual perception (Lamme & Roelfsema, 2000; VanRullen & Thorpe, 2001) and neuroscience (Boly et al., 2011; Koch et al., 2016).

# 5. Conclusion

Which neural network architecture matches human behavior in an artificial grammar learning task? We demonstrate in this study that recurrent neural network models are closer to humans than feedforward ones, irrespective of the grammars' level in the Chomsky's hierarchy. This result endorses recurrent models as the more plausible cognitive architecture underlying language, pointing to an essential role of recursion in grammar learning (Rohrmeier, Dienes, Guo, & Fu, 2014). However, it could be fetching to generalize this conclusion to natural language acquisition (Corballis, 2014; Jackendoff, 2011; Fitch & Friederici, 2012; Vergauwen, 2014): the Chomsky hierarchy, despite being an excellent reference, does not embrace all natural grammars (Jäger & Rogers, 2012), and does not fully reflect human cognitive complexity (Öttl et al., 2015). Additional studies will investigate more ecological grammars to further corroborate our conclusion (Smith, 2003).

## Acknowledgments

This work was funded by an ERC Consolidator Grant P-CYCLES number 614244. We are grateful to Jacob Martin for fruitful discussion about the Chomsky's hierarchy. Parts of this work were done while RV was visiting the Simons Institute for the Theory of Computing.

# Appendix A

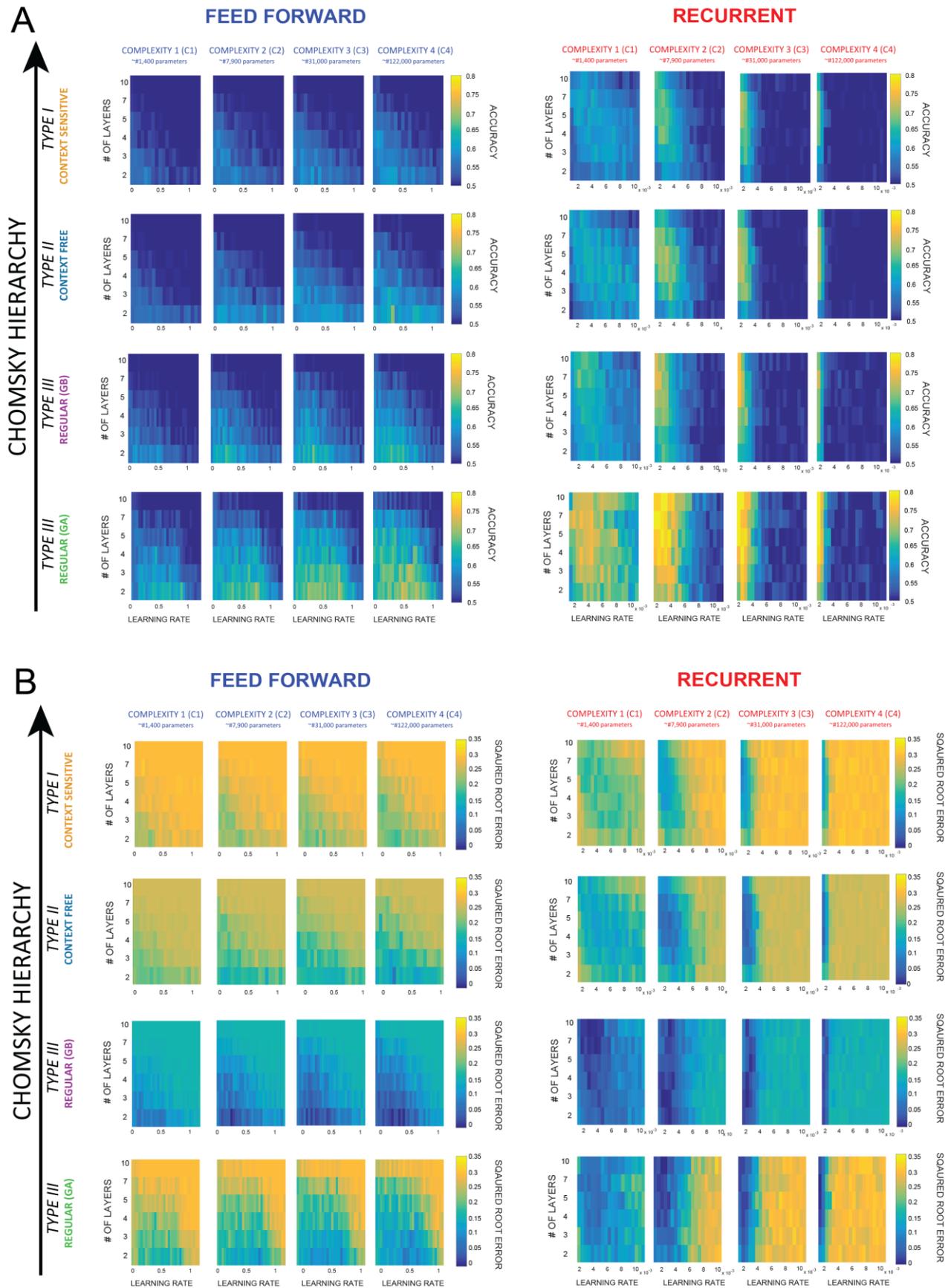

**Figure A.1**. Results of the parameter search for each complexity levels (columns, increasing number of parameters from left to right). Rows represent grammars ordered according to the Chomsky's hierarchy. Subplot A shows the accuracy for feedforward (left) and recurrent (right) networks. Subplot B shows the squared root error, i.e. the distance to human performance. Non-

surprisingly, the performance improves with more parameters, but similar patterns of results are observed between complexity levels: networks closer to human behavior have 2 layers in FF architectures, and lower learning rates in RR ones.

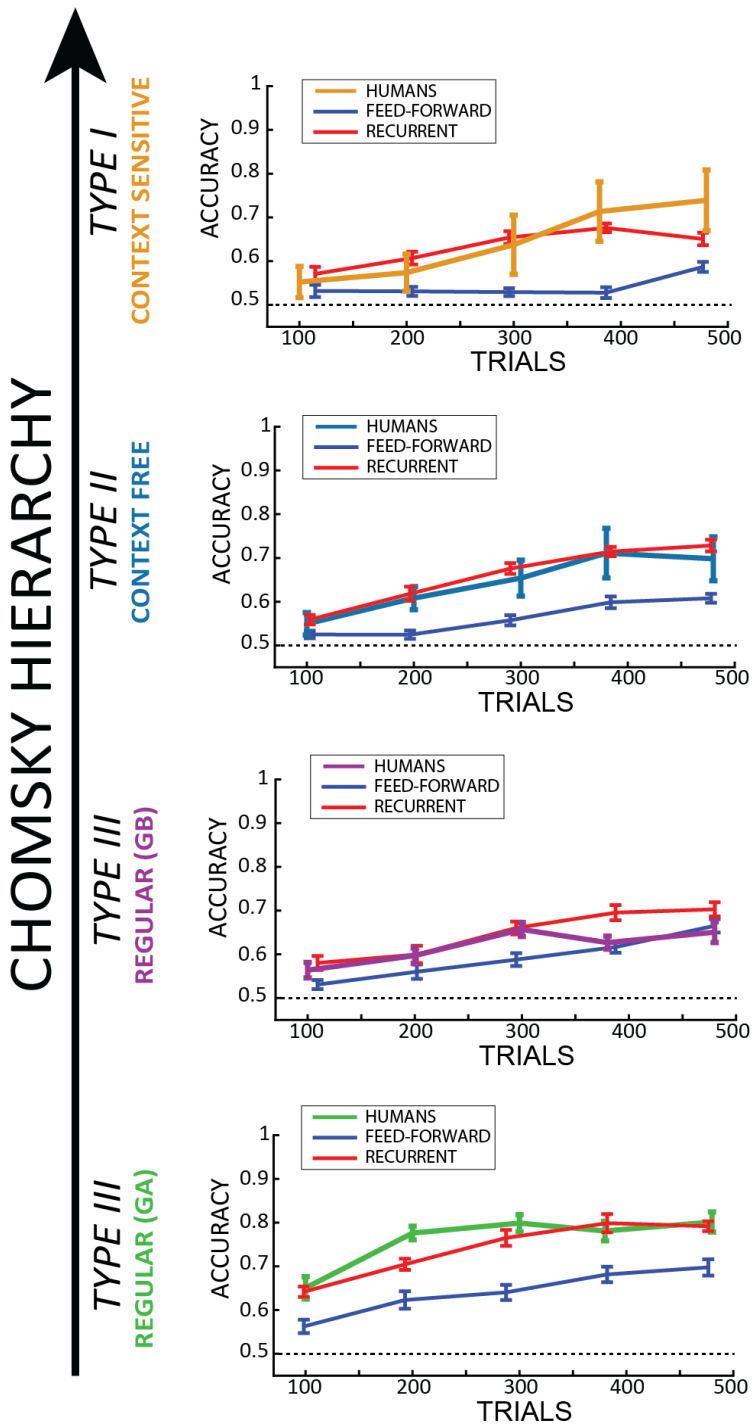

**Figure A.2**. The figure show the results of the FF and RR model which obtained the performance closest to human when averaging over all grammars (see figure 3 C,D). Notice that in each subplot, the FF and RR networks are the same across grammars. As shown in figure 4, the results confirm that RR architectures are closer to humans than FF, except in grammar B.

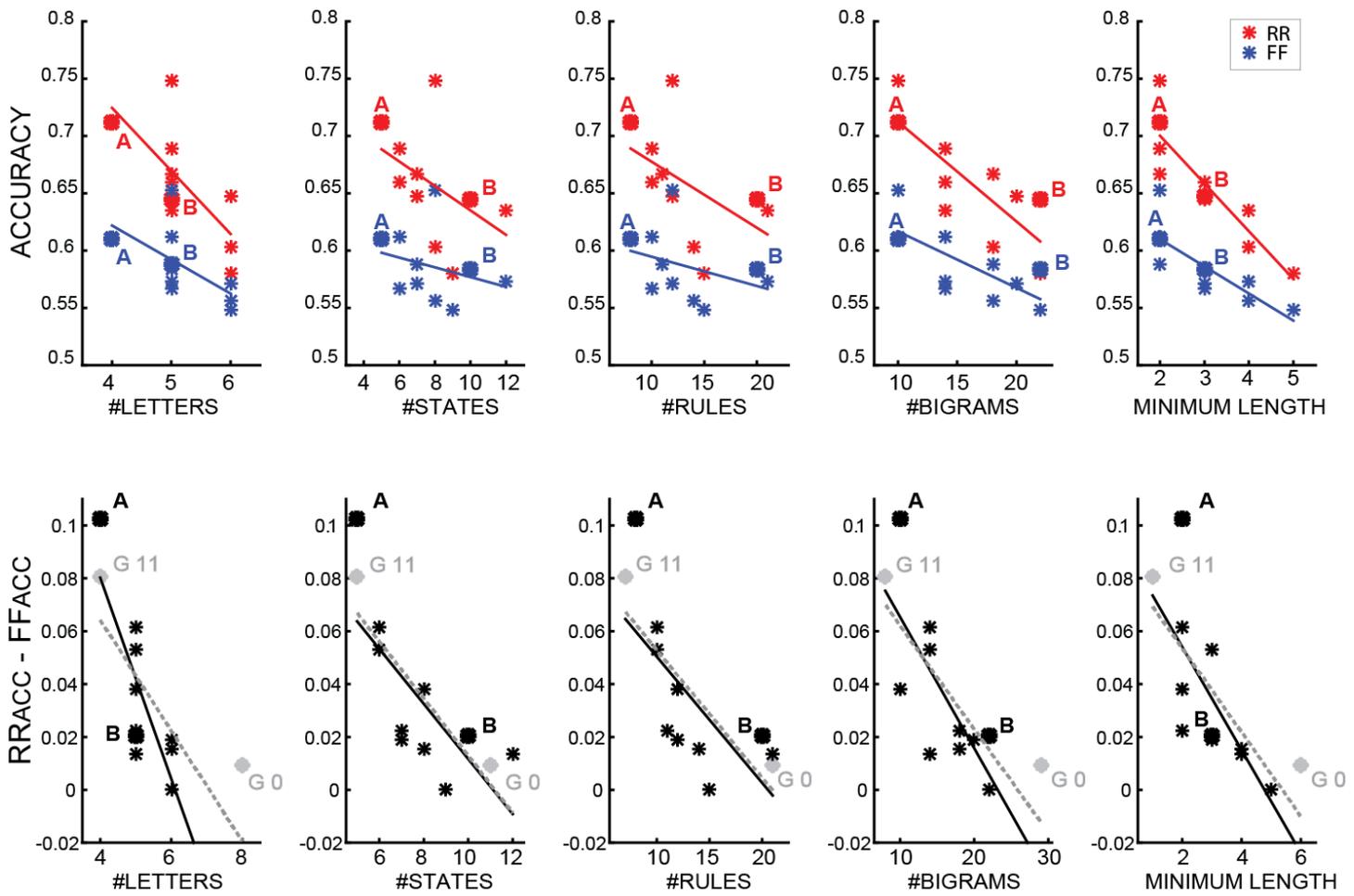

**Figure A.3**. Upper panel: correlation between the accuracy of FF (in blue) and RR (in red) networks with the different metrics used to evaluate grammar complexity. Grammar A and B are highlighted in each plot. Lower panel: same correlations but with the difference between RR and FF performance. Grammars A, B and grammars 0 and 11 are highlighted in each subplot.

# Appendix B

**Table B.1** – Results of the Bayesian ANOVA testing accuracy, comparing groups and the last three blocks. The factor AGENT is composed of FF, RR and Humans. The label '8' refers to the last block of the implicit session, 'M' and 'E' refers respectively to the memory and explicit blocks. Each cell is composed of a BayesFactor and the percentage error in the form: BF(%err).

| Grammars | AGENT | 8 vs E | 8 vs M | E vs M |
|---|---|---|---|---|
| GA | 102.11 (0.009) | 889.7 (7e-7) | 1.2 (0.002) | 2.5 (8e-4) |
| GB | 1.22 e7 (7e7) | 1.1e6 (1e-10) | 0.4 (1e-4) | 7272.2 (9e-9) |
| CFG | 2.28 (0.008) | 3.96 (4e-4) | 1.6 (1e-5) | 0.35 (4e-5) |
| CSG | 7.44 (0.006) | 3.85 (1e-4) | 2.75 (1e-4) | 0.39 (0.005) |
| CFG (impl) | 18.37 (0.011) | 20.3 (5e-4) | 5.7 (4e-4) | 0.38 (0.005) |
| CSG (impl) | 10.11 (5e-4) | 4.71 (1e-5) | 3.5 (4e-4) | 0.4 (4e-6) |

**Table B.2** – Results of the Bayesian ANOVA testing accuracy over AGENT (FF, RR and Humans, with post hoc comparisons), SEQUENCE LENGTH and their interaction. Each cell is composed of a BayesFactor and the percentage error in the form: BF(%err).

| Grammars | AGENT | SEQ. LENGTH | INTERACTION | H vs FF | H vs RR | RR vs FF |
|---|---|---|---|---|---|---|
| GA | 4.9e8 (0.8) | 1.2e56 (0.3) | 18.47(0.8%) | 4.5e7 (2e-14) | 0.26 (3e-5) | 6.9e7 (2e-13) |
| GB | 0.38 (0.022) | 6.5e26 (0.008) | 3.3e3(1.1%) | - | - | - |
| CFG (impl) | 1.2e5 (0.020) | 7.1e6 (3e-4) | 8.3e10(0.8%) | 3.2e3 (4e-7) | 0.21 (0.02) | 2.0e5 (1e-9) |
| CSG (impl) | 2.0e5 (0.010) | 1.2e8 (1e-4) | 5.8e10(1.2%) | 46 (1e-4) | 1.1 (0.004) | 4.8e5 (2e-8) |

**Table B.3** – Results of the Bayesian correlation between the FF-RR difference in performance and the complexity metrics in regular grammars. Further than the Pearson's r index and the Bayes Factor, also 95% confidence intervals of the 'r' estimate are reported in the last column.

| Complexity metrics: | Pearson's r | Bayes Factor | Upper – Lower 95% CI |
|---|---|---|---|
| #LETTER | -0.796 | 20.834 | [-0.23 -0.93] |
| #RULES | -0.676 | 5.764 | [-0.10 -0.88] |
| #BIGRAM | -0.732 | 9.704 | [-0.15 -0.91] |
| #MIN.LEN | -0.678 | 5.823 | [-0.08 -0.87] |
| #STATE | -0.721 | 8.667 | [-0.14 -0.90] |

# Appendix C

**Questionnaires C.1:** questionnaire provided to the participants after the regular grammar B, to test the level of rules' awareness. A very similar questionnaire was provided after grammar A.

SbjNo :                          Gender :                          Age :

# De-briefing questionnaire

*Please answer each question accordingly to the sequences you have been classifying during the experiment*

1) Which letter(s) is(are) more likely to be in the first position

   M     R     T     V     X          how confidence you feel about your response
                                      0 (no confident at all) – 100 (very confident)

2) Which letter(s) is(are) more likely to be in the last position

   M     R     T     V     X          how confidence you feel about your response
                                      0 (no confident at all) – 100 (very confident)

3) Which letter(s) is(are) more likely to be in the second position

   M     R     T     V     X          how confidence you feel about your response
                                      0 (no confident at all) – 100 (very confident)

4) Which letter(s) cannot be presented twice consequently (e.g. in position 2 and 3, or in position 3 and 4, etc)

   M     R     T     V     X          how confidence you feel about your response
                                      0 (no confident at all) – 100 (very confident)

5) Which letter(s) is(are) more likely to appear after the letter 'X'

   M     R     T     V     X          how confidence you feel about your response
                                      0 (no confident at all) – 100 (very confident)

6) Which letter(s) is(are) more likely to appear after the bigram 'MX'

   M     R     T     V     X          how confidence you feel about your response
                                      0 (no confident at all) – 100 (very confident)

7) Which letter(s) is(are) more likely to appear after the bigram 'XT'

   M     R     T     V     X          how confidence you feel about your response
                                      0 (no confident at all) – 100 (very confident)

**Questionnaires C.2:** questionnaire provided to the participants after the context-specific and context-free grammars to test the level of rules' awareness.

SbjNo :                              Gender :                              Age :

# De-briefing questionnaire

Please answer each question accordingly to the sequences you have been classifying during the experiment

1. Here a list of sequences that do not comply with the rules:
   - can you report (circle / underline) the error in each of them? (report your confidence about your response: from 0 to 100)
   - can you write the correct version of each sequence ? (report your confidence about your response: from 0 to 100)

| Sequence : | Correct sequence + confidence: |
|---|---|
| SEQUENCE I | |
| SEQUENCE II | |
| .. | |
| SEQUENCE VII | |
| | |
| | |
| | |
| | |
| | |

2. Which pattern(s) or rule(s) have you been following to perform the task?